\documentclass[a4paper]{article}
\usepackage{graphicx}
\usepackage{amsmath}
\usepackage{amsfonts}
\usepackage{amssymb}
\usepackage{bbm}
\usepackage{xcolor}
\usepackage{authblk}

\title{Towards scalable pattern-based optimization for dense linear algebra}

\author[1]{D\'aniel Ber\'enyi*}

\author[2]{Andr\'as Leitereg}

\author[2]{G\'abor Lehel}

\affil[1]{GPU Lab, Wigner Research Centre for Physics, Budapest, Hungary}
\affil[2]{Faculty of Informatics, E\"otv\"os Lor\'and University, Budapest, Hungary}
\date{}
\begin{document}
\maketitle





\abstract{Linear algebraic expressions are the essence of many computationally intensive problems, including scientific simulations and machine learning applications.
However, translating high-level formulations of these expressions to efficient machine-level representations is far from trivial: developers should be assisted by automatic optimization tools so that they can focus their attention on high-level problems, rather than low-level details.
The tractability of these optimizations is highly dependent on the choice of the primitive constructs in terms of which the computations are to be expressed.
In this work we propose to describe operations on multi-dimensional arrays using a selection of higher-order functions, inspired by functional programming, and we present rewrite rules for these such that they can be automatically optimized for modern hierarchical and heterogeneous architectures. Using this formalism we systematically construct and analyse different subdivisions and permutations of the dense matrix multiplication problem.}


\maketitle

\section{Introduction}\label{sec1}

The computational problems we use computers to solve grow ever more complex and demanding, and we need appropriately powerful tools to help us tackle them.
While the hardware which performs these computations is advancing at a sufficient rate to keep pace with our increasing demands, the programming languages and development tools we use to implement them are not, resulting in increasing burdens being placed on the shoulders of programmers.
A profound example of this can be seen in the optimization of large linear algebraic expressions that arise in scientific simulations, multi-dimensional data processing (such as image- and video analysis), and machine learning applications.
Naively written algorithms tend to greatly underutilize the capabilities of modern hardware.
The cause of this, at root, is in the hierarchical memory and cache architectures and multi-level parallelism that characterizes such hardware.
At a low level, most computations are limited by memory bandwidth: it is difficult to consistently keep the execution hardware supplied with new data for it to work on.
In many cases, the same data is accessed repeatedly, and storing it far away from the processor results in excessive transfers, high latency, and a waste of bandwidth.
Unfortunately, fast memory which is close to the processor is expensive in terms of area and cooling, and therefore available only in limited quantities.
Together, these considerations give rise to hierarchical cache- and memory layouts where large memories are slow and fast memories are small, and software must take care to place data carefully in order to maximise utilization.

Common levels of the hierarchy are the following:

\begin{itemize}
	\item SIMD (Single Instruction Multiple Data) parallelism, and registers
	\item Thread level parallelism, and thread-private memory
	\item Thread-group level parallelism, and thread-local memory (GPU) or caches (CPU), shared within, but not between groups
	\item Device level parallelism, and device memory (within an entire CPU or GPU)
	\item Node level parallelism, involving multiple devices (CPUs and GPUs) which communicate via an internal bus (can have shared address space)
	\item Cluster level parallelism, involving nodes which communicate via a network (data must be serialized)
\end{itemize}

At the lowest level, which is the level of instructions, hardware uses vector instructions which can process 2, 4, 8, 16, or even more pieces of data simultaneously.
Instructions are executed concurrently in multiple threads, transferring memory between their private memory and registers.
Modern CPUs often support tens of concurrent threads, and GPUs support thousands.
At the next level up, a single computational node often contains many separate CPUs and GPUs.
Finally, multiple computational nodes are connected together using a network to form clusters and grids.
To efficiently perform a computation, it must be split into progressively smaller parts at each level of the hierarchy, but one would prefer not to have to do this splitting manually in each case, because they are in many regards extremely similar: at a high level of abstraction, splitting, for example, a summation of many elements between vector instructions and between nodes of a cluster requires analogous logic, only a separate implementation.

The traditional approach to linear algebra is BLAS (Basic Linear Algebra Subroutines, \cite{BLAS}), which provides hand-optimized versions of many of the most common operations, such as scalar-vector, vector-vector, vector-matrix and matrix-matrix routines, for single and double precision floating point as well as complex number types.
There are two problems with this approach.
The first is that if a required operation is not in the collection, either due to differing structure, or to operating on a different number type, then the pre-selected collection of operations provides no tools with which to build it.
The second, that calls to these predefined routines are not efficiently composable: intermediate results of one operation must be written out to memory before the next one can use them, resulting in wasteful transfers.
As a complementary approach, much effort has been spent on optimizing loops in existing imperative code bases, as done, for example, by the Polly optimizer \cite{Polly}.
Unfortunately, as the structure of loops is very flexible, it is highly difficult to implement sufficiently powerful pattern recognition to find optimization opportunities in hand-written code.
Newer approaches (such as Eigen \cite{Eigen}, or Armadillo \cite{Armadillo} and other competing libraries) first construct an expression tree out of the supplied high-level primitives, analyze it, and only then attempt to fuse operations within it.

A similar situation arises in machine learning, especially when done using neural networks, where the user constructs high-level graphs describing the computation out of predefined primitives.
For example, the widely used machine learning framework TensorFlow \cite{Tensorflow} suffers from the same forced memory write-out problem as BLAS, and only recently is an optimizing just-in-time compiler (XLA \cite{XLA}) being developed which attempts to fuse and optimize operations in the computational graph in order to eliminate temporaries.

In every case, the root of the problem is twofold: what are the best primitives to use in the domain specific language in order to describe the operations the users want to represent? And what are the patterns (particular combinations of these primitives) which can be automatically detected and recombined in order to yield a more optimal execution, both in terms of memory usage as well as parallelism?

Most approaches to solving these problems naturally divide them into multiple levels: a higher level consisting of (say) algorithmic primitives, a lower level of hardware-specific building blocks, and sometimes, even further lower-level representations beyond that \cite{Lift}\cite{OpenCLRewriteCodegen}.
In this work, we primarily focus on the higher level problem of choosing the right primitives in a theoretically justified way and constructing their associated rewrite rules in such a way that the result is a closed system, rather than delving into the details of the lower-level primitives, which both differ at each of the different levels of parallelism mentioned above, and have also already been extensively studied by others \cite{OpenCLRewriteCodegenMatMul}\cite{LiftIL1}.
Our sense is that in the existing approaches, a significant amount of genericity has been missed in the choice of the high-level building blocks, and that this limits their scalability.

In this paper, we proceed as follows.
First, we provide some examples of fusion problems from linear algebra and neural networks.
Next, we assemble a collection of primitives which together are flexible enough to express a wide range of useful operations, and which are highly amenable to optimization.
Further on, we describe rewrite rules which can be automatically applied to an expression tree consisting of these primitives in order to fuse or rearrange operations within it. Finally, we demonstrate on the matrix multiplication problem how these rewrite rules can improve data locality and performance.

\section{Motivating examples}\label{sec2}

Inflexible libraries centered around providing pre-written, non-parametric primitives suffer from too many temporaries, for example the expressions:

\begin{equation}
w_i = \sum_j (A_{ij} + B_{ij}) \cdot (v_j + u_j),
\end{equation}

\begin{equation}
C_{ik} = \sum_j A_{ij} \cdot B_{jk} \cdot g_j,
\end{equation}
can be fused into a minimal modifications of the usual matrix-vector and matrix-matrix multiplication formulas respectively, but one wants to do such transformations automatically.
Obviously the terms can contain other more composite but still low arithmetic density operations (scalar multiplications: $y_i = c \cdot x_i$, outer products: $A_{ij} = x_i \cdot y_j$).
One cannot expect that a single primitive can encompass this variability, so analysis is required on the whole subexpression tree.
In the case of neural networks, layers are usually composed of some dense arithmetic transformation (affine transformation or convolution), a non-linear function and some normalization in between \cite{BatchNormalization}:

\begin{equation}
y_k^b = \sum_i W_{ik} x_i^b + \beta_k,
\label{nn_dense}
\end{equation}

\begin{equation}
z_k = \frac{y_k^b - E[y^b]}{ \sqrt{ V[y^b] } },
\label{nn_bn}
\end{equation}

\begin{equation}
r_k = h(z_k)
\label{nn_nl}
\end{equation}
where eq. \ref{nn_dense}. is a dense transformation, that can be a convolution too, eq. \ref{nn_bn}. is batch normalization, where E is the average, V is the variance over the collection indexed by k, and finally eq. \ref{nn_bn} is an element wise applied non-linearity.

Here, again, it is preferable, to fuse all these steps into a single operation without temporaries, because the last two parts are of low arithmetic density.
While common cases can be hand written for end users, due to the fast pace of developments in the field modifications of all these steps and the invention of new ones pose a challenge to framework creators.
Another important problem is concerned with properly dividing the data of a computation into the hierarchical memory layout of the target device.
Again, we might strive to do this for arbitrary expressions, rather than for just pre-written primitives.
Consider the earlier example, or a slight variation of it:

\begin{equation}
C_{ik} = \sum_j A_{ij} \cdot B_{jk} \cdot g_j
\end{equation}

\begin{equation}
C_{ipq} = \sum_{jk} A_{ijk} \cdot B_{jp} \cdot C_{kq} \cdot g_j \cdot f_k
\end{equation}
Such formulas pop up in the numerical solution of partial differential equations, and they directly follow from the structure of the original equation.
The important question here is what the optimal partitioning of the data in the hierarchical memory is to maximize data reuse (caching) and thus maximize performance.

Both linear algebra and neural networks deal with multidimensional data: scientific computations usually deal with 2 or 3 dimensions but for some applications higher dimensions are also common.
Neural networks also usually deal with 2 or 3 dimensional data (images or video), but these have color dimensions also.
When processing images or frames usually multiple samples are grouped together (called batching) that creates another dimension.
Altogether, these languages need primitives that can abstract over dimensionality, and be parameterizable so future extensions would not necessarily need new primitives.
In the following, we build a language suitable for expressing multi-dimensional linear algebra operations, such as vector, matrix or tensor products and later construct rewrite rules that can optimize the expressions built in this language to be efficient on modern hardware endowed with hierarchical memory and parallelism.

\subsection{Functional DSL for dense linear algebraic expressions}

Most dense operators that are common in linear algebra, image processing or in neural networks are different kinds of repetitive operations on a composite structure, where the structure might be multidimensional.
The operations can be reductions, dimension keeping transformations, or enlarging operations.
Simplification and fusion of these operations is of high interest, because these form the innermost and most extensively used parts of many modern computational solutions, from Big Data to Machine Learning or large scale simulations.
Traditional imperative formulations (like for loops) are way too general to be suitable for straightforward optimization, or require complex and expensive analysis before that. Functional primitives, that are more restricted, however can provide suitable expressivity, easier automatization and parallelisation possibilities.
Here, we assume that all the dimension, shape and layout information is represented at the type level (and thus, known at compile time), although we may omit them to simplify the notation.
A multidimensional flat array type is represented as $a^{(e_1, e_2, ... e_{dim})}$, with the constructor:

\begin{equation}
\mathtt{data \ Array_0 :: (dim :: Nat) \rightarrow (e_1, e_2, ..., e_{dim}) \rightarrow * \rightarrow * }
\end{equation}
where $dim$ is the number of dimensions, and $e_k$ is the extent in each dimension.
Such representations were investigated earlier \cite{JGibbonsAPL} and Naperian functors were found to be a useful abstraction, which essentially state that such containers are identical to functions from an index set to the elements:

\begin{equation}
\mathtt{f \ a \simeq Idx \ n \rightarrow a}
\end{equation}
An important property is that composition of Naperian functors can be indexed by product of indices:
\begin{equation}
\mathtt{f \ (g \ a) \simeq Idx \ n_f \times Idx \ n_g \rightarrow a}
\end{equation}
and product of Naperian functors can be indexed by the sum of indices:

\begin{equation}
\mathtt{(f \ a, \ g \ a) \simeq Idx \ n_f \ |\  Idx \ n_g \rightarrow a}
\end{equation}
Naperian functors were named after John Napier, the inventor of the logarithm, due to the close resemblance to the logarithmic identities $\log (f \circ g) \simeq \log f \times \log g$ and $\log (f \times g) \simeq \log f + \log g$.
However, to us a more important property is that nested Naperian functors can be transposed: $f \hspace{1mm} (g \hspace{1mm} a) \simeq g \hspace{1mm} (f \hspace{1mm} a)$ (see \cite{JGibbonsAPL} for details).
In our case we not just simply use this property for expressing the transposition of functor structures, but rather to express exchange of subdivisions over such containers as demonstrated below.

Nested, or subdivided arrays (arrays of arrays) can be represented by iterating the exponentiation,

\begin{eqnarray}
&&\mathtt{data \ Array ::} \notag \\ 
&&\mathtt{(dim :: Nat, \ depth :: Nat) \rightarrow  ( (e_1, e_2, ... e_{dim}), (f_1, ..., f_{dim}), ..._{depth}) \rightarrow * \rightarrow * } \notag \\
\end{eqnarray}
for example: $a^{(2, 3)(4, 5)}$ is a matrix of elements of type $a$, such that the outer level is divided into 4 block rows and 5 block columns, and each of these sub blocks is a matrix with 2 rows and 3 columns, so the whole structure contains $(2 \times 4 \cdot 3 \times 5) = 120$ elements.

However, as we are going to use operations that consume strictly one (the outermost) dimension of the input structure, it is easier to formulate the computation if we represent the subdivided arrays as flat multidimensional ones. For this we use strides (step sizes over the data corresponding to each dimension) denoted as $a^{((e_1, s_1), (e_2, s_2),...,(e_n, s_n))}$. For the above 120 elements (stored in a linear memory) $a^{((3,1), (2,3), (5,6), (4,30))}$ is a simple 4 dimensional tensor in row-major order, but if we interpret the same elements as $a^{((3,1), (2,15), (5,3), (4,30))}$, then we get the described subdivided array. As we are to change the number and the order of the dimension consuming operations, we need to make corresponding changes in the logical layout of the data:

\begin{itemize}
	
	\item \texttt{\textbf{subdiv d b s}}: creates a new subdivision of structure $s$ by splitting the extent at dimension $d$ into blocks of size $b$. 
	\newline If $subdiv \ d \ b \ (M :: a^{(e_0, s_0) ... (e_{n-1}, s_{n-1})}) = M' :: a^{(e_0', s_0') ... (e_{n}', s_{n}')}$ then

	$\begin{cases}
    (e_i', s_i') = (e_i, s_i), \ (i<d)\\
    (e_d', s_d') = (b, s_d), \ {\rm where \ b \ must \ be \ a \ divisor \ of \ }e_d\\
    (e_{d+1}', s_{d+1}') = (e_d/b, b \cdot s_d), \\\
    (e_i', s_i') = (e_{i-1}, s_{i-1}), \ (i>d+1).
    \end{cases}$

	\item \texttt{\textbf{flatten d s}}: merges the adjacent depths $d$ and $d+1$ into a single layer, it is the inverse of $subdiv$. If $flatten \ d \ (M :: a^{(e_0, s_0) ... (e_{n}, s_{n})}) = M' :: a^{(e_0', s_0') ... (e_{n-1}', s_{n-1}'})$ then
	
	$\begin{cases}
	(e_i', s_i') = (e_i, s_i), \ (i<d)\\
	(e_d', s_d') = (e_d \cdot e_{d+1}, s_d), \ \\
	(e_{i-1}', s_{i-1}') = (e_{i+1}, s_{i+1}), \ (i>d)
	\end{cases}$
	
	\item \texttt{\textbf{flip $d_1$ $d_2$ s}}: flips (transposes) the layout by swapping any two dimensions (extent and stride together). flip applied twice is identity and it is commutative in $d_1$ and $d_2$. If we omit the second argument, its default value is $d_2 = d_1 + 1$.
	
\end{itemize}

To compare the representation and the functions with \cite{JGibbonsAPL}, we see that this model can represent arbitrary, but same dimensional subdivision and collapse of a given multi dimensional array.
This follows the intuition of practical problems, that subdivide the initial structure into sub chunks while keeping the dimension but divide the extents to improve memory usage.
The extra constraint, that we have between the extents is that they must properly divide the number of elements in the given direction, and thus we cannot simply say, that we have a list (composition) of arbitrary functors in the container type, but we need to express the specific numbers as list of tuples and introduce the above operations with regards to these extents and dimensions.

We note that flip really corresponds to the flip for functions:

\begin{equation}
\mathtt{flip :: (a \rightarrow b\rightarrow c)\rightarrow(b\rightarrow a\rightarrow c)}
\end{equation}
due to the Naperian function connection.
It seems tempting to associate \texttt{subdiv} with \texttt{curry} and \texttt{flatten} with \texttt{uncurry}, but we have the extra constraint of divisibility.

We now list the three most common operations on the above introduced arrays:

\begin{equation}
\mathtt{map :: (a\rightarrow b) \rightarrow f \ a \rightarrow f \ b}
\end{equation}

\begin{equation}
\mathtt{zip :: (a \rightarrow b \rightarrow c) \rightarrow  f \ a \rightarrow f \ b \rightarrow f \ c}
\end{equation}

\begin{equation}
\mathtt{reduce :: (a\rightarrow a \rightarrow a) \rightarrow f \ a \rightarrow a}
\end{equation}
\texttt{map} and \texttt{zip} (in Haskell: \texttt{zipWith}) are elementwise operations over an array f: these can represent e.g. scalar multiplication or division of vectors, matrices and tensors, or colour transformations of pixels of an image, or in the case of zip elementwise sums, differences.
\texttt{reduce} as its name suggests can reduce a structure into a summary value using a binary operator: this can be used to calculate sums, products, minimal/maximal values etc. In contrast to \texttt{fold}, it takes at least 1 element.
Another important property of these operations is that they are straightforward to parallelize.
map and zip are considered to apply their function argument completely independently for each element, or element pairs.
For reduce, if the binary operation is associative, we can regroup the reduction and if it is also commutative we can even reorder it.
These transformations are important for optimizing memory access patterns \cite{CUDAReduce}.

We can now translate formulas to our DSL, for example, consider the vector-matrix product:
\begin{equation}
u_i = \sum_j A_{ij} \cdot v_j
\end{equation}

\begin{equation}
\mathtt{
	u = map (\backslash r \rightarrow reduce \ (+) \ (zip \ (*) \ r \ V) \  A
}
\end{equation}
where $A$ is a row-wise matrix of $n$ elements per row and $v$ is an array of $n$ scalars.
The main property we want to optimize is the number of times new threads are spawned.
Each higher order function may represent multiple parallel threads, which are costly to spawn.
Also, each of these calls may need temporary memory allocated to store intermediate results.
Minimizing both of these drastically improves performance on modern hardware.

\section{Rewrite rules for higher-order functions}\label{sec3}

The fundamental idea of the approach presented here is that the possible rewrites are induced by the structure of types and operations on them, ie.: given a simple expression of a computation as a combination of the above primitives, we would like to automatically change it in such a way that it produces the same results but with different (probably better) runtime performance and more suitability for parallel execution.
This cannot be purely driven by the types of the arrays, because their relations are expressed in the higher-order functions operating on them, however types can express the matching subdivision structures and track rearrangements and signal potential mistakes in the implementation of the rewrites.

The important question is: can we identify a set of rewrite rules which together suffice to optimize any possible combination of the above primitives?

We consider two groups of rewrite rules:

\begin{enumerate}
	\item Cases where the higher-order functions form a pipeline (sequential composition), with the output of one being the input of the next. These are be fused by composition operations.
	\item Cases where one higher-order function is passed to another as an argument function for it to apply.
	These cases describe operations on nested structures. We cannot fuse them, but we are interested in exchanging them implicitly relying on the Naperian functor property.
\end{enumerate}

To present the rewrite rules we use Haskell syntax, but we emphasize that these rules follow from simple operations on functions (since these types are Naperian functors) that can be represented in virtually any language.

In the first group we start with the simplest fusion law.
From the functor property, we know that:

\begin{equation}
\mathtt{
	map \ f \ . \ map \ g = map \ (f \ . \ g)
}
\end{equation}
This allows us to collapse any length of composed maps into a single call to map.
The question is, whether we can do similarly for zip or zip-map compositions.
It turns out that the composition of other zips before the arguments of a zip is not closed as it can lead to three or more parameters to the zip, so we need to generalize them to arbitrary number of arguments.
Let us now define the n-ary version of map and zip that we call nzip:

\begin{equation}
\mathtt{
	nzip :: (a_n \rightarrow b) \rightarrow (f \  a)_n \rightarrow f \ b
}
\end{equation}
where the notation $a_n$ is understood as:

\begin{equation}
\mathtt{
	a_0 \rightarrow a_1 \rightarrow a_2 \rightarrow ... \rightarrow a_n
}
\end{equation}
for any finite n. We may reduce the definition if we wish to:

\begin{equation}
\mathtt{
	nzip :: (a_n) \rightarrow (f \ a)_n.
}
\end{equation}
Now, if we define the generalised composition:

\begin{eqnarray}
& \mathtt{ncomp} & \mathtt{:: (i :: Nat) \rightarrow (a_n \rightarrow c) \rightarrow (b_m \rightarrow a_i)} \notag \\	
& & \mathtt{\rightarrow} \mathtt{(a_{0 ... i-1} \rightarrow b_0 \rightarrow ... \rightarrow b_{m-1} \rightarrow a_{i+1 ... n-1} \rightarrow c)}   \\
& \mathtt{ncomp} & \mathtt{ \_ \ f \ g = ...}
\end{eqnarray}
that composes g before the i-th argument of f (we index arguments from 0), and we can state the general composition rule for nzip as (see also Figure \ref{nzipfusion}):

\begin{figure}
	\includegraphics[width=0.95\textwidth]{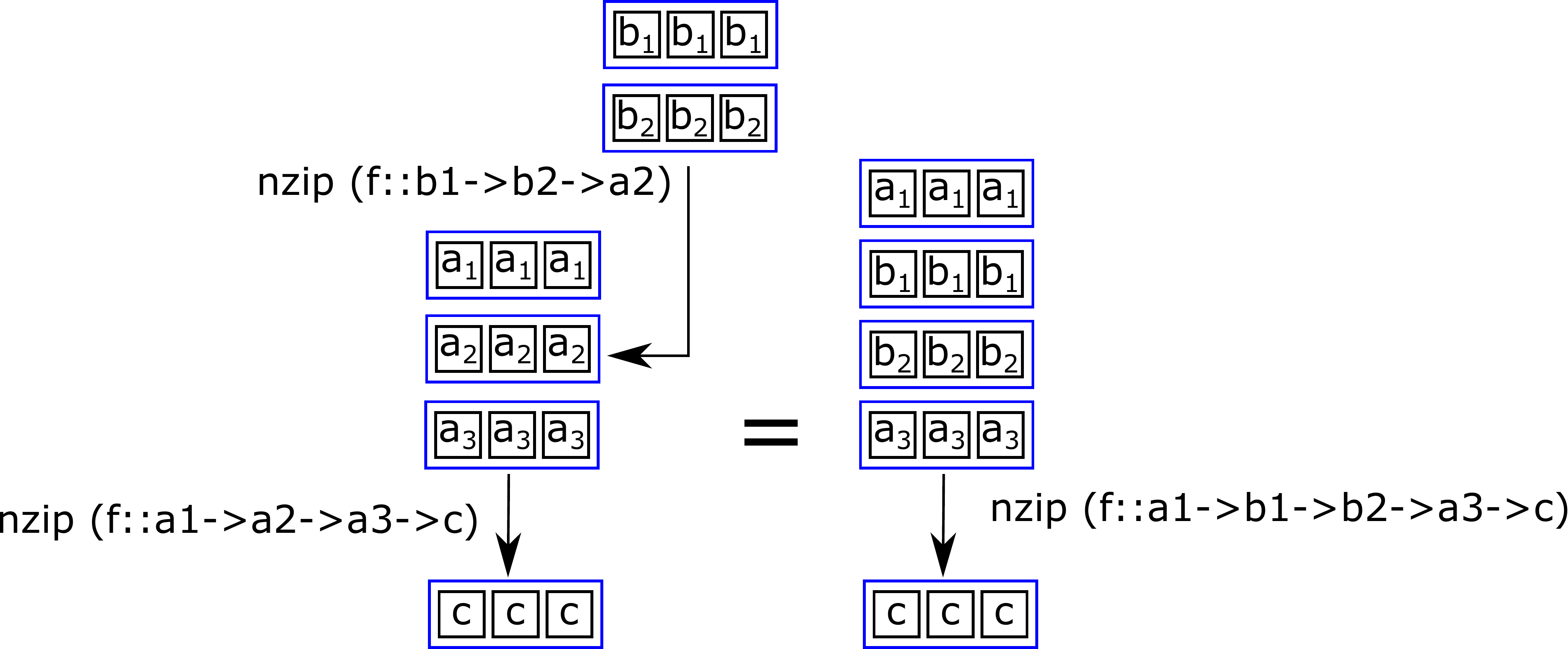}
	\caption{Illustration of nzip fusion}
	\label{nzipfusion}
\end{figure}

\begin{eqnarray}
&&\mathtt{ nzip \ f \ xs_{0 ... i-1} \ (nzip \ g \ ys) \ xs_{i+1 ... n-1} =} \notag \\
&&\mathtt{ nzip \ (ncomp \ i \ f \ g) \ xs_{0 ... i-1} \ ys_{0 ... m-1} \ xs_{i+1 ... n-1}}
\end{eqnarray}
This way nzip (and thus map and zip) is now closed under arbitrary compositions.
Next we turn to reductions.
It seems feasible to separate the reduction function from all preprocessing/merging done on the arguments, so we define reduce-of-zips (following the idea of foldMap from the Haskell library Data.Foldable) for arbitrary number of arguments:

\begin{equation}
\mathtt{
	rnz :: (b\rightarrow b \rightarrow b) \rightarrow (a_n \rightarrow b) \rightarrow (f \ a)_n \rightarrow b
}
\end{equation}
The first argument is a reduction function, that is assumed to be at least associative, the second argument is a zipping function that elementwise zips the final n arguments of rnz.
Now, we can consider the problem of composing arbitrary maps or zips before a reduction and fusing them into a single reduction, for a map:

\begin{equation}
\mathtt{
	rnz \ r \ f \ (nzip \ g \ xs_{0 ... n-1}) = rnz \ r \ (ncomp \ 0 \ f \ g) \ xs_{0 ... n-1}
}
\end{equation}
or for a zip:

\begin{eqnarray}
&&\mathtt{ rnz \ r \ f \ (nzip \ g \ xs_{0 ... n-1}) \ (nzip \ h \ ys_{0 ... n-1}) } = \notag \\
&&\mathtt{ rnz \ r \ (ncomp \ 0 \ (ncomp \ 1 \ f \ g) \ h) \ xs_{0 ... n-1} \ ys_{0 ... n-1} }
\end{eqnarray}
One important application is that now we can express the dot product of vectors: $c = \sum_i u_i \cdot v_i$ in a single call as:

\begin{equation}
\mathtt{
	dot \ u \ v = reduce \ (+) \ (zip \ (*) \ u \ v) = rnz \ (+) \ (*) \ u \ v
}
\end{equation}
In the general case, reductions with different first arguments cannot be fused.
Even if r is the same, the outer map function f should be identity so that the whole construction expresses the regrouping of a single reduction and thus can be fused to one.

For completeness, we review some rules concerning products.
The implicit assumption here is that products of types are usually used to represent a smaller amount of data than arrays, despite that arrays are just repeated products of the same type.
One can express the infamous Array-of-Structures vs. Structures-of-Arrays optimization pattern as:
\begin{equation}
\mathtt{
	Array \ dim \ layout \ (a, \ b) = (Array \ dim \ layout \ a, \ Array \ dim \ layout \ b)
}
\end{equation}
The simplest rule concerning the higher-order functions is
\begin{equation}
\mathtt{
	(map \ f \ x, \ map \ g \ y) = map \ (f, \ g) \ (x, \ y)
}
\end{equation}
Where $(f, g)$ is understood as the function product ($\mathtt{(***)}$ in Haskell's Control.Arrow) $\mathtt{:: (a\rightarrow b)\rightarrow(c\rightarrow d)\rightarrow (a, \ c)\rightarrow (b, \ d)}$. If the argument is the same for both functions, we can make use of the fanOut function $\mathtt{ (fanOut::(a\rightarrow b)\rightarrow(a\rightarrow c)\rightarrow a\rightarrow(b, \ c))}$:
\begin{equation}
\mathtt{
	(map \ f, \ map \ g) \ x = map \ (fanOut \ f \ g) \ x
}
\end{equation}
Trivially we can extend these definitions to arbitrary number of products, and also extend them for the other functional primitives:
\begin{equation}
\mathtt{
	(zip \ f \ x \ y, \ zip \ g \ p \ q) = zip \ (f, \ g) \ (x, \ y) \ (p, \ q) 
}
\end{equation}
\begin{equation}
\mathtt{
	(reduce \ f \ x, \ reduce \ g \ y) = reduce \ (f, \ g) \ (x, \ y)
}
\end{equation}
with the generalization of the product to multiple arguments: \newline$\mathtt{(a \rightarrow b \rightarrow c)\rightarrow (x\rightarrow y \rightarrow z) \rightarrow (a, x)\rightarrow(b, y)\rightarrow(c, z)}$  and assuming that the functors are all of compatible sizes.

We now turn to cases that are not compositions, but are operating on nested structures, such that the higher-order-functions are nested in their function parameters.
We are interested in structure changing rules that can exchange the order of higher-order functions so we can use them to regroup computations.

The simplest such rewrite is flipping maps. Consider the dyadic product:
\begin{equation}
A_{ij} = v_i \cdot u_j
\end{equation}
this can be written as:
\begin{equation}
\mathtt{
	map \ (\backslash x \rightarrow map \ (\backslash y \rightarrow x*y) \ u) \ v
}
\end{equation}
or
\begin{equation}
\mathtt{
	map \ (\backslash y \rightarrow map \ (\backslash x \rightarrow x*y) \ v) \ u
}
\end{equation}
up to a flip in the functor structure: the first result matrix is expressed as an array of rows, while the second as columns. Note, that the multiplication is done in the same order so commutativity is not required for this transformation to work. It is straightforward to extend this transformation to zips, so this is omitted here.
The less trivial transformation is, that maps and reductions can be flipped in a similar way, consider the matrix-vector product:
\begin{equation}
v_i = \sum_j A_{ij} \cdot  u_j.
\end{equation}
If $A$ is a standard flat multidimensional array in row-major order, then the multiplication can be expressed in the textbook way as:
\begin{equation}
\mathtt{
	map \ (\backslash r \rightarrow rnz \ (+) \ (*) \ r \ u) \ A
} \label{eq:matmul1}
\end{equation}
that can be also written as:
\begin{equation}
\mathtt{
	rnz \ (zip \ (+)) \ (\backslash c \ q \rightarrow map \ (\backslash e \rightarrow e*q) \ c) \ (flip \ 0 \ A) \ u \label{eq:matmul2}
}
\end{equation}
(we pulled out the rnz part from the definition of dot). The former is performing the computation by taking the dot product of each row and the vector, while the latter corresponds to the multiplication of the columns of the matrix by elements of the vector and adding them together with the vector addition.
The crucial point here is, that the selected value from the vector (q) is reused for the whole column, at the cost that more memory is needed to store the accumulator for the reduction, but more parallelism can be exploited. Before we can formalize the exchange operation, we first introduce $\mathtt{lift}$ that is the applicative operation raising a function to operate over functors (shorthand for a partially applied map):
\begin{equation}
\mathtt{
	lift :: (a\rightarrow b) \rightarrow f \ a \rightarrow f \ b
}
\end{equation}
Of course, we can generalize the above identities to n-ary arguments and can operate at any level of the functor nestings.
\begin{figure}
	\includegraphics[width=0.95\textwidth]{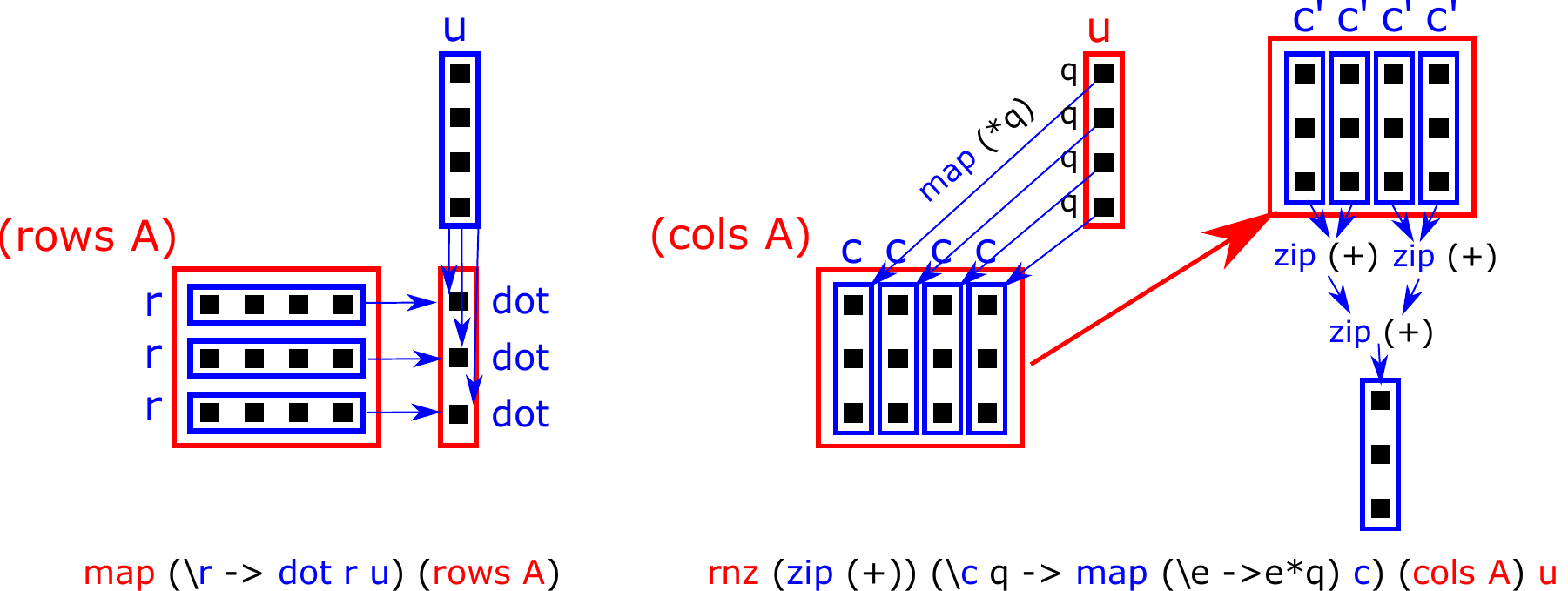}
	\caption{Two versions of matrix-vector multiplication related by the flip identity (eq. \ref{flip_identity})	for map and rnz operations.}
	\label{matmulvct1}
\end{figure}
The formal version of the exchange rule used in the matrix-vector product above is (see Figure \ref{matmulvct1}):
\begin{eqnarray}
&&\mathtt{ map \ (\backslash a \rightarrow rnz \ r \ m \ a \ u) \ A =} \notag \\
&&\mathtt{ rnz \ (lift \ r) \ (\backslash a \ q \rightarrow map \ (\backslash \alpha \rightarrow m \ \alpha \ q) \ a) \ (flip \ 0 \ A) \ u
\label{flip_identity}
}
\end{eqnarray}
This transformation is one of the most important transformations that can generate cache friendly versions of a given formula.
Looking at the type level, we find, that the matrix $a^{(n,m)}$ is flipped (resulting in $a^{(m,n)}$) to keep the same dimension binding to the structure of the vector. So we can conclude, that exchanging two nested higher order functions must be done with an appropriate flip in the subdivision structure.
It is also possible to state the flip identity generally for rnz operations, but it is much more restricted, the reduction operation must be the same and must be commutative and associative:
\begin{eqnarray}
&& \mathtt{ rnz } \  \mathtt{ r \ (\backslash a_1 \ a_2 \rightarrow rnz \ r \ m \ a_1 \ a_2 \ B) \ A_1 \ A_2 =  } \\
&& \mathtt{ rnz } \ \mathtt{ r \ (\backslash a_1 \ a_2 \ b \rightarrow rnz \ r \ (\backslash \alpha_1 \ \alpha_2 \rightarrow m \ \alpha_1 \ \alpha_2 \ b) \ a_1 \ a_2) \ (flip \ 0 \ A_1) \ (flip \ 0 \ A_2) \ B}. \notag
\end{eqnarray}
The idea of the proof of this transformation is based on expanding the $\mathtt{ rnz }$ operations and observing that all $(m \ \alpha_1 \ \alpha_2 \ b)$ elements appear exactly once on both sides and then the equality follows from the properties of $r$.

These operations are necessary, but not sufficient to bring competitive performance when compared to hand-written functions.
The reason is, as mentioned earlier, the hierarchical memory layout of all modern computing hardware.
For example, in a well written matrix multiplication code, the matrix is subdivided into smaller submatrices usually as many times as many levels are in the memory hierarchy: for example, GPUs have global memory, shared (local) memory and vector registers.

We can construct identities because subdivisions should not alter the results of operations when compared to the execution without subdivisions.
As a trivial example, mapping over a vector can be done by mapping the same function over the subdivisions, and then this whole operation over the outer structure, or even over repeated subdivisions:
\begin{eqnarray}
&\mathtt{map}& \mathtt{f \ v} = \notag \\
&\mathtt{map}& \mathtt{(\backslash x \rightarrow map \ f \ x) \ (subdiv \ d \ b \ v) } = \notag \\
&\mathtt{map}& \mathtt{(\backslash x \rightarrow map (\backslash y \rightarrow map \ f \ y) \ x) \ (subdiv \ d_1 \ b_1 (subdiv \ d_2 \ b_2 \ v)) } \notag \\
\end{eqnarray}
The same is true for zips and reductions.
The important observation is, that all the actual computation is done in the innermost map, all outer maps are just "logical" operations, reshapings.
Not only are operations independent of the level of subdivisions, but by invoking the earlier identities, we can rearrange their order.

We now use the above results to regroup the calculation of the matrix-vector product, where A has the type $a^{(n, m)}$:
\begin{equation}
v_i = \sum_j A_{ij} \cdot  u_j
\end{equation}
\begin{equation}
\mathtt{
	map \ (\backslash r \rightarrow dot \ r \ u) \ A \label{eq:matmulvct1_1}
}
\end{equation}
subdividing the rows r and the vector u into smaller chunks of size $b$, we can write:
\begin{eqnarray}
\mathtt{
	map \ (\backslash r \rightarrow rnz \ (+) \ (\backslash b \ c \rightarrow dot \ b \ c) \ r \ (subdiv \ 0 \ b \ u)) \ A^\prime } \notag \\
\mathtt{ A^\prime = subdiv \ 1 \ b \ A } \label{eq:matmulvct1_2}
\end{eqnarray}
On the other hand, starting from the other formula:
\begin{equation}
\mathtt{
	rnz \ (zip \ (+)) \ (\backslash c \ q \rightarrow map \ (\backslash e \rightarrow e*q) \ c) \ (flip \ 0 \ A) \ u \label{eq:matmulvct2_1}
}
\end{equation}
and subdividing the columns of the matrix we find:
\begin{eqnarray}
\mathtt{
	rnz \ (zip \ (+)) \ (\backslash C \ c \rightarrow map \ (\backslash r \rightarrow dot \ r \ c) \ C) \ A^{\prime\prime} \ (subdiv \ 0 \ b \ u) } \notag \\
\mathtt{ A^{\prime\prime} =  subdiv \ 1 \ b (flip \ 0 \ A) } \label{eq:matmulvct2_2}
\end{eqnarray}
\begin{figure}
	\includegraphics[width=0.99\textwidth]{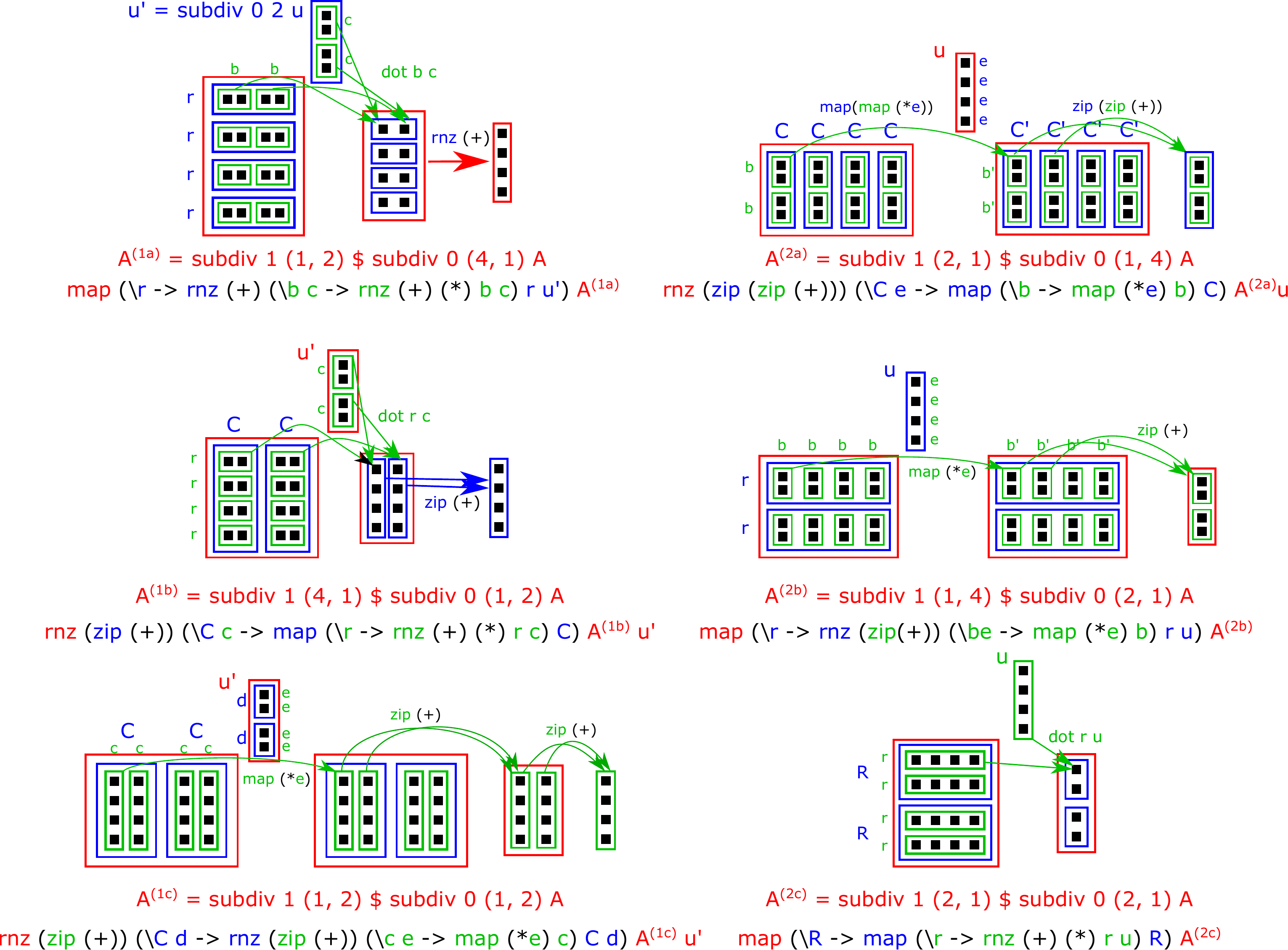}
	\caption{Six rearrangements of the matrix-vector multiplication gained by subdividing (eq. \ref{eq:matmulvct1_1}) (left three cases) and (eq. \ref{eq:matmulvct2_1}) (right three cases) and then applying the map-rnz flipping rule of eq. \ref{flip_identity}.}
	\label{matmulvct2}
\end{figure}
Comparing the two results, we can see that it is still the manifestation of the map-reduce flip identity constructed earlier, so subdivisions and flips commute as expected.
One can construct other arrangements by applying the map-rnz flip formula above, here we list all six cases (see also the Figure \ref{matmulvct2}):
\begin{itemize}
	\item 1a: $\mathtt{ map \ (\backslash r \rightarrow rnz \ (+) (\backslash b \ c \rightarrow rnz \ (+) \ (*) \ b \ c) \ r \ u^\prime) \ A^{(1a)} }$  \newline \hspace*{\fill} (eq. \ref{eq:matmulvct1_2}),
	\item 1b: $\mathtt{ rnz \ (zip \ (+)) \ (\backslash C \ c \rightarrow map \ (\backslash r \ rnz \ (+) \ (*) \ r \ c) \ A^{(1b)} \ u^\prime }$  \newline \hspace*{\fill}   (applying flip 0 to 1a)
	\item 1c: $\mathtt{ rnz \ (zip \ (+)) (\backslash C \ d \rightarrow rnz \ (zip \ (+)) (\backslash c \ e \rightarrow map \ (*e) \ c) \ C \ d) \ A^{(1c)} \ u^\prime }$ \newline \hspace*{\fill}  (applying flip 1 to 1b)
	
	\item 2a: $\mathtt{ rnz \ (zip \ (zip \ (+))) \ (\backslash C \ e \rightarrow map \ (\backslash b \rightarrow map \ (*e) \ b) \ C) \ A^{(2a)} \ u}$   \newline \hspace*{\fill} (applying subdiv to map in eq. \ref{eq:matmulvct2_1} ),
	\item 2b: $\mathtt{ map \ (\backslash r \rightarrow rnz \ (zip \ (+)) (\backslash b \ e \rightarrow map \ (*e) \ b) \ r \ u ) \ A^{(2b)}}$           \newline \hspace*{\fill}   (applying flip 0 to 2a)
	\item 2c: $\mathtt{ map \ (\backslash R \rightarrow map \ (\backslash r \rightarrow rnz \ (+) \ (*) \ r \ u ) \ R ) \ A^{(2c)}}$             \newline \hspace*{\fill}   (applying flip 1 to 2b)
\end{itemize}
with the $A^{(..)}$s being matching subdivisions, $\mathtt{u^{\prime} = subdiv \ 0 \ b \ u}$.

Here we write out how the different matrices are related to the original one (c.f. Figure \ref{matmulvct2}):

\begin{itemize}

	\item $\mathtt{A^{(1a)} = subdiv \ 0 \ 2 \ A}$
	\item $\mathtt{A^{(2a)} = subdiv \ 0 \ 2 \ (flip \ 0 \ A)}$
	\item $\mathtt{A^{(1b)} = flip \ 1 \ (subdiv \ 0 \ 2 \ A)}$
	\item $\mathtt{A^{(2b)} = flip \ 1 \ (subdiv \ 0 \ 2 \ (flip \ 0 \ A))}$
	\item $\mathtt{A^{(1c)} = flip \ 0 \ (flip \ 1 \ (subdiv \ 0 \ 2 \ A))}$
	\item $\mathtt{A^{(2c)} = flip \ 0 \ (flip \ 1 \ (subdiv \ 0 \ 2 \ (flip \ 0 \ A)))}$
	
\end{itemize}
In cases 1a 1b 1c the vector is subdivided (we divided the rnz operation that involves the vector), in the other three cases the vector is not subdivided as we subdivided the map that does not involve the vector.

Comparing the formulas, we once again see that the exchange of the higher-order functions happens simultaneously with the flipping of the logical structure.
Furthermore, the formulations listed using the same letter (for example, 1a and 2a) differ only in a full ("deep") transposition of the logical structure.
Each representation provides different parallelization and caching possibilities, but may also require different memory capacities for accumulation.
While the final performance is affected by many parameters, a few important points can be observed:
If the vector is subdivided (1a, b, c), then it is possible to have 2-level caching (with one block of the vector at an outer level, and one element at an inner level); while if it is not (2a, b, c), then only a single level of the memory hierarchy can be used.
Cache efficiency is improved when reductions involving the vector are propagated outwards, thereby reusing the blocks or components multiple times.
This, on the other hand, causes the size of temporaries used for the reductions to become bigger, for example, 1a uses only scalar accumulators, while 1b and 1c require full columns.
This can form a limit on how high the reductions can be raised.
If the matrix is stored in a row-major order, then the memory access patterns of the 1c and 2a variants may be too interleaved, and similarly for 1a and 2c if using column-major storage.
Repeated subdivisions of the same dimension (1c, 2c) provide no real benefit over the original formula.

\section{Demonstrative results}\label{sec4}

In this section, we show some brief results to illustrate, that subdividing and rearranging a computation expressed with the above higher-order functions can indeed yield better performing versions. For this demonstration we consider matrix multiplication for square matrices on CPUs. We implemented\cite{DataView} a DSL in C++ that transforms an AST containing the above introduced HoFs and lambda abstraction / application nodes. We implemented the pattern match and replace logic by structured recursion schemes (catamorphisms, anamorphisms, paramorphisms and Elgot-algebras), and defined the pattern-replacement pairs for standard lambda calculus transformations ($\eta$ equivalence, $\beta$ reduction) and the subdivision and flip operations as defined above. The system can take an AST of a simple computation as input, can perform subdivision and then can generate all permutations of HoFs nested under each other as seen below. Since this kind of nesting forms a list, the well known Steinhaus-Johnson-Trotter algorithm \cite{Trotter}\cite{Selmer} can be used to enumerate all possible permutations by adjacent element swapping.

Finally, for each permutation we can generate C++14 code for a small library defining the HoF implementations and Views for accessing the data. The Views store their dimensions and strides in template parameters thus implementing the Naperian functor concept (as they have a fixed shape). Flips and subdivisions are implemented as template functions. Below all results are for 1024 x 1024 double precision matrices measured on a Core i5 7300HQ. We also implemented the naive algorithm in C as well as a block-optimized version by hand, which ran in 4.9 seconds and 0.278 seconds, respectively.
\begin{table}
	\centering
	\begin{tabular}{ | l  l  l | r |}
		\hline
		\multicolumn{3}{|c|}{HoF permutation} & Time [s] \\ \hline
		mapA & rnz  & mapB & 0.45 \\ \hline
		rnz  & mapA & mapB & 1.41 \\ \hline
		mapA & mapB & rnz  & 4.67 \\ \hline
		mapB  & mapA & rnz & 6.05 \\ \hline
		rnz  & mapB & mapA & 13.8 \\ \hline
		mapB  & rnz & mapA & 15.6 \\ \hline
	\end{tabular}
	\caption{Six rearrangements of the naive matrix-matrix multiplication. The HoF order from left to right is the nesting from top down in the computation. The naive C level implementation is 4.9 seconds and the improved blocked version is 0.278 seconds. }
	\label{matmul_lvl3}
\end{table}
\begin{table}
	\centering
	\begin{tabular}{ | l l l l | r |}
		\hline
		\multicolumn{4}{|c|}{HoF permutation} & Time [ms] \\ \hline
		rnz  & mapA & mapB & rnz  & 186  \\ \hline
		mapA & rnz  & mapB & rnz  & 249  \\ \hline
		rnz  & mapA & rnz  & mapB & 324  \\ \hline
		mapA & rnz  & rnz  & mapB & 478  \\ \hline
		rnz  & rnz  & mapA & mapB & 1468  \\ \hline
		mapB & rnz  & mapA & rnz  & 2024  \\ \hline
		rnz  & mapB & mapA & rnz  & 2231  \\ \hline
		mapA & mapB & rnz  & rnz  & 4970  \\ \hline
		rnz  & mapB & rnz  & mapA & 5580  \\ \hline
		mapB & mapA & rnz  & rnz  & 5686  \\ \hline
		mapB & rnz  & rnz  & mapA & 5972  \\ \hline
		rnz  & rnz  & mapB & mapA & 7353  \\ \hline
	\end{tabular}
	\caption{Twelve rearrangements of the matrix-matrix multiplication, where the $rnz$ operation is subdivided. The HoF order from left to right is the nesting from top down in the computation. The naive C level implementation is 4.9 seconds and the improved blocked version is 278 ms. }
	\label{matmul_lvl4}
\end{table}

\noindent Let us start from the mathematical form:
\begin{equation}
C_{ik} = \sum_j A_{ij} B_{jk}
\end{equation}
The equivalent naive functional formulation is:
\begin{equation}
\mathtt{
	C = map \ (\backslash r_A \rightarrow map \ ( \backslash c_B \rightarrow rnz \ (+) \ (*) \ r_A \ c_B) \ B) \ A \label{eq:matmul_1}
}
\end{equation}
where $r_A$ are the rows of $A$, and $c_B$ are the columns of $B$, we assume that they are both stored in a row major format, so accessing their consecutive elements have different costs. While this textbook formulation is easy to understand and implement, its performance is far from optimal due to inefficient cache usage. Using the exchange rules above, we can create 6 permutations (since each HoF is different) and measure their performance (Table \ref{matmul_lvl3}). 

It is easy to understand why the best performance is achieved when mapB is in the innermost position: in this case a flip was used to exchange the rnz with mapB (compared to the original form) and thus matrix B is accessed in the inner loop row-wise, in the same direction as the row of A in the rnz, and such consecutive reads are the best for the memory controller logic. The opposite order leads to the most suboptimal case, when B is outside and A is inside: in this case both of them are accessed column wise, which effectively invalidates most of the caches.

Next, we may subdivide the $rnz$ part, and create all permutations of the 4 HoFs, however we do not differentiate between the two $rnz$s, so effectively we have 12 different cases (Table \ref{matmul_lvl4}). At the subdivision we use $b=16$ blocksize.

\begin{figure}
	\centering
	\includegraphics[width=0.45\textwidth]{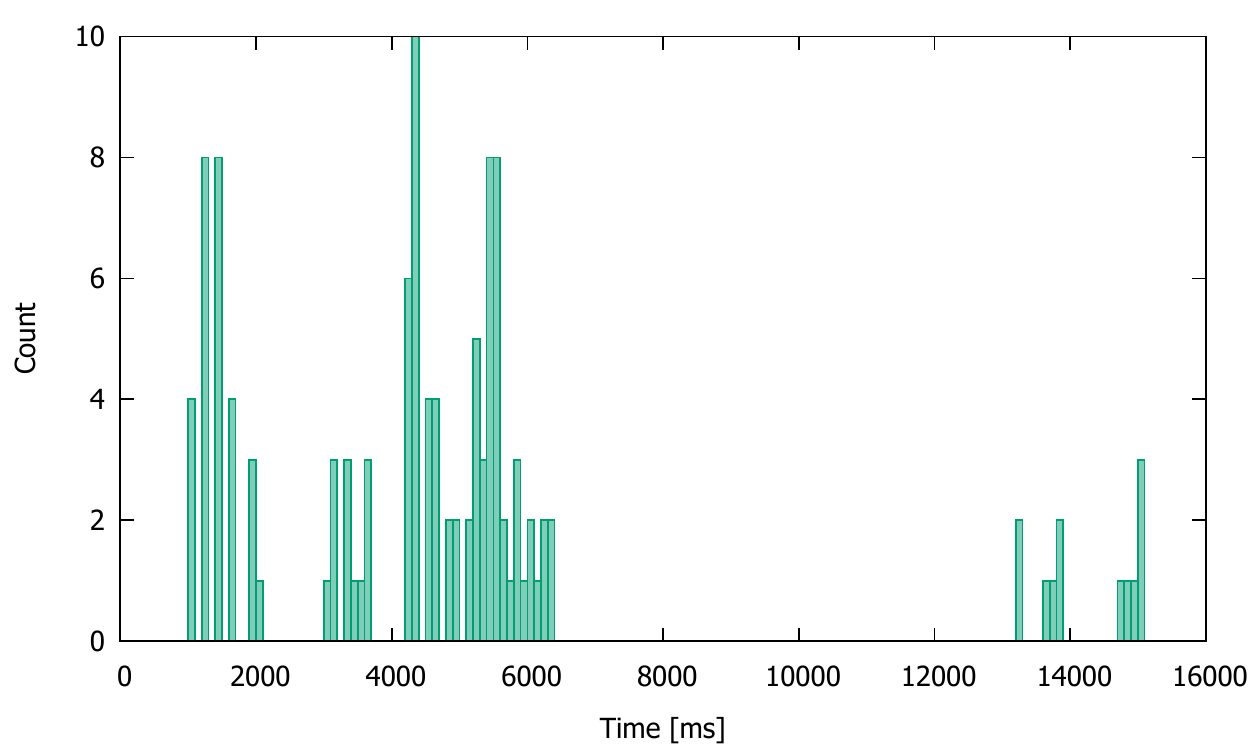}
	\caption{Rearrangements of the matrix-matrix multiplication gained by subdividing the two map operations.}
	\label{matmul_lvl5b}
\end{figure}

\begin{figure}
	\centering
	\includegraphics[width=0.45\textwidth]{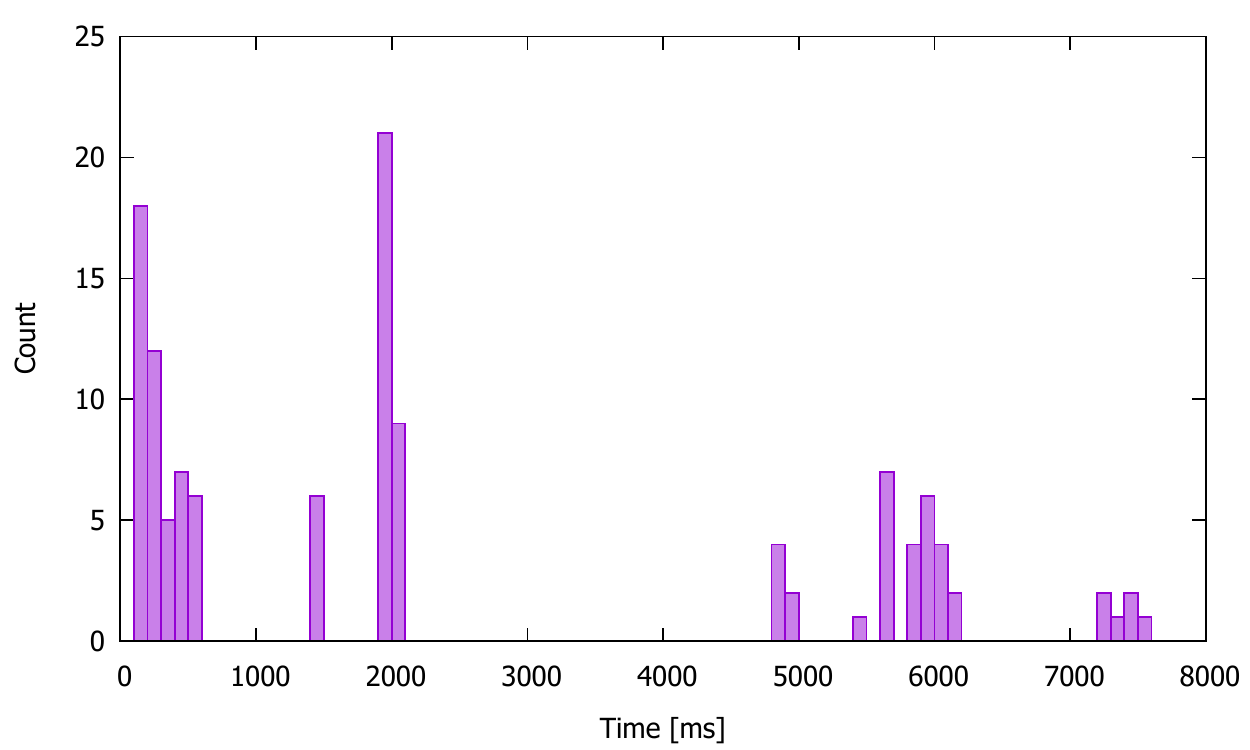}
	\caption{Rearrangements of the matrix-matrix multiplication gained by subdividing the rnz operation twice.}
	\label{matmul_lvl5}
\end{figure}

\begin{figure}
	\centering
	\includegraphics[width=0.45\textwidth]{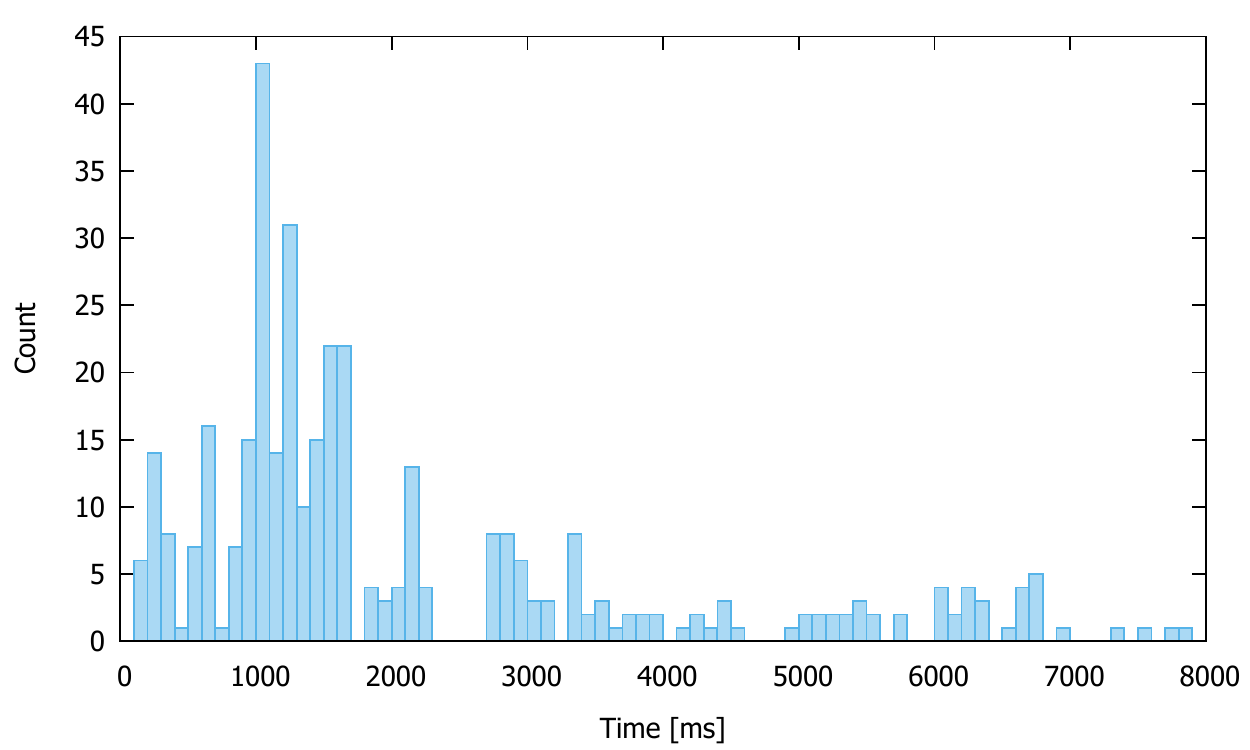}
	\caption{Rearrangements of the matrix-matrix multiplication gained by subdividing all HoFs once.}
	\label{matmul_lvl6}
\end{figure}

In the other case, when the two $map$s are subdivided (in their outermost direction) does not improve performance: the best candidate there needs 1 second (c.f. Figure \ref{matmul_lvl5b}). When the $rnz$ is subdivided twice, the average performance is drastically increased as shown in Figure \ref{matmul_lvl5}, all candidates are at least as good as the naive implementation, but the best candidate is just as good as in Table \ref{matmul_lvl4}. Subdividing both the $map$s and the $rnz$ yields no improvement over the case where just the $rnz$ was subdivided (Figure \ref{matmul_lvl6}). 

We note, that Eigen with vectorization explicitly disabled (but the compiler may still vectorize) performs the matrix multiplication in 333 ms, and with its own optimal vectorization it finishes in 60 ms. These results of course depend not only on the hardware at hand, but also on the compiler used (in our case clang SVN r323406), and especially in our compile-time sized case, on the size of the matrix. The important point is, that with the automatic rewrites it was possible to achieve more than 25 fold increase in performance (4.9 seconds to 180 ms) w.r.t. the naive implementation.

On GPUs, not all potential rearrangements are feasible. To make use of the 2 dimensional thread spawning capability, it is useful to select cases, where the maps over the matrices are adjacent. This is immediately the case in the naive form, and from the subdivided cases we compare with the one arising from the simultaneous subdivision of all three HoFs, namely: mapA mapB rnz mapA mapB rnz. With the necessary additional considerations of local memory, this leads to a 40\% improvement (AMD Radeon HD7970, ComputeCpp 0.4 compiler).

By applying the presented rewrite rules repeatedly, we get deeper subdivisions and further rearrangements of the initial formula, and the additional degrees of freedom can be used to utilize vectorization both on CPUs and GPUs, or over multiple devices and threads, but this is outside the scope of this paper and left to future work

\section{Related work}\label{sec5}

In this work we focused on designing a collection of functional primitives, which are composable and closed under composition as often as possible, and suitable for pattern based optimisation to increase data locality. Here, we review other approaches to utilizing hardware parallelism.

\emph{Algorithmic Skeletons} or algorithmic patterns are higher-order functions that aim to capture some recurring structures of computation, especially in the context of parallel programming \cite{ColeThesis}\cite{SkeletonHoFs}\cite{SkeletonHoFTransformations}. Such a collection of algorithms usually includes divide-and-conquer for recursively formulated computations, iterative combination for graph problems, and also pipelining (a sequence of stages, where different stages on different objects are executed in parallel), farming (a variant of map) and other algorithms. Most of these skeletons were formed by abstracting common high-level patterns of problems (even higher-level, than the ones discussed in this work) and while their formulation is similar to higher-order functions in functional programming, their theoretical background, compositionality and interplay with other language constructs like products and sums types is not self-evident. Nevertheless, these approaches provide patterns which we know how to execute optimally on parallel hardware.

\emph{Optimization and fusion in the context of function programming using rewrite-rules} is extensively researched for improving languages and compilers, like Haskell \cite{StreamFusion}, the GHC compiler \cite{HaskellDataFlowFusion, GHCRewrite} and Data-Parallel Haskell\cite{DataParallelHaskell}. Most of the efforts are centered on eliminating the needless construction and consumption of lists and streams in memory, which are frequent in such languages. While fusion of such generic data structures is important, further optimization is possible in special cases, where fusion would require additional constraints on the shape of data, as is the case with arrays and especially arrays of arrays (multidimensional arrays). For example, a list of lists may be irregular (each contained list may have a different length, potentially zero), and thus may not be safely transposed (order of nestings exchanged), whereas a fixed size array of fixed size arrays must be regular, making such a transformation possible. This limitation was recognized, and fusion for arrays in particular has also been investigated, as in the case of \cite{Chakravarty}. SAC\cite{SAC} is another functional array library with an emphasis on fusion. In the HPC setting, rewrites of legacy code based on user annotations are described in \cite{HPCRewrite}. The Spiral language \cite{SPIRAL} that targets performance critical problems with extensive inlining and optimizations is also driven by rewrite rules. Delite \cite{Delite} also uses rewrite rules to optimize locality in NUMA code, and targets CUDA capable GPUs.

\emph{GPU Computing} is traditionally done through an imperative API (CUDA, OpenCL, SYCL, OpenACC, OpenMP) and/or an abstraction library over them, like Copperhead\cite{Copperhead} (data parallel constructs embedded in Python) or Accelerator\cite{Accelerator} (parallel arrays with predefined operations in C\#).

From the functional perspective Obsidian\cite{Obsidian} and Accelerate\cite{Accelerate1, Accelerate2} employ techniques similar to those presented here, but none of these involve the optimization of operations on multidimensional arrays. The NOVA language shares some of the fusion logic presented here \cite{NOVA} and supports nested structures, but nested fold-map structures are flattened. Multidimensional and nested parallel constructs are discussed in \cite{Lee}, but they do not change the layout of the computation, only the mapping of nested loops to GPU threads. Loop exchange in the context of small linear algebraic problems was studied in \cite{Spampinato}, but not in the general case of arbitrary tensors of nested structures. 

\vspace{0.5cm}

We mention two projects in more detail as they use very similar methodology to the one presented here.

\emph{Lift} \cite{Lift}\cite{LiftThesis}\cite{LiftIL1} is a DSL for high-level GPU programming, it generates OpenCL code from high-level functional primitives with rewrite rules. A restricted set of primitives are provided which can be nested, opening the possibility of utilizing nested parallelism on the hardware. While the map, zip and reduce constructs are similar to the ones presented here, they are not generalized to the variadic case, preferring to express the same thing using nested applications instead, but this results in a larger number of potential program rearrangements which need to be assessed. Map subdivision (which the authors call splitting) is aided by auxiliary functions (split and join), whereas we are able to nest and transform HoFs directly.  During rewriting, they explore multiple different algorithmic choices (by applying different rewrite rules) to generate multiple different optimizations of the input program, which can then be lowered to OpenCL-specific parallel primitives. Their rewrite rules does not involve the exchange rules presented here, although some of their features can be expressed using their ${\mathtt reorderstride}$ function. Overall, for linear algebra problems, their approach may require a larger number of primitives, and exploring a larger search space, compared to the more generic constructs presented here.

\emph{Futhark} \cite{FutharkThesis}\cite{FutharkProc} is another purely functional DSL which, similarly to Lift, targets GPUs. It has a wide range of generic primitives for operating on arrays which are similar to the ones presented here and come with an extensive set of fusion rules. The Futhark multi-stage optimising compiler also uses rewrites to apply optimizations. The fusion rules are selected to never duplicate computations and never reduce available parallelism. Futhark uses a data flow graph reduction logic and, because evaluating fusion of arbitrarily large graphs is NP-hard, they use a greedy algorithm which selects the first suitable candidate. As for exchange rules, Futhark performs generic map-loop, map-reduce interchanges and avoids full flattening, but may end up reducing parallelism during these transformations in an attempt to keep data locality. The exchange and transformation logic is predefined and specialized for GPUs, and might not be well-suited for other architectures. In the case where there is more than one candidate for the most efficient rearrangement, the Futhark compiler generates multiversioned code and selects between them at runtime. After performing rearrangements, the kernel extraction pass selects parts of the program which will be lowered to be executed on the GPU. Shape of arrays is manipulated by index space transformations and explicit rearrangements in memory that are inserted by the compiler after it performs complicated analysis of access patterns. In our approach, we choose the opposite direction, that is, generating different traversal patterns as a consequence of changes in the logical structure of the data. Tiling in Futhark is likewise depends on access analysis, and is restricted to one or two dimensions. Whereas in our approach, tiling also falls out as a consequence of rearranging HoFs and the logical structure of the arrays they act upon and does not have any inherent limitations in the number of dimensions it supports. In the SGEMM benchmark of Futhark, the authors note that further improvements would be possible by using register based tiling in addition to local memory tiling. Both of these may be expressed in our approach by mapping different layers of the nested HoF structure to different memory spaces, as in our discussion of the memory hierarchy in the introduction.

\section{Conclusions}\label{sec6}

We have investigated a collection of rewrite rules for the fusion, subdivision and rearrangement of higher-order functions expressing dense array arithmetic, which are relevant to the optimization of linear algebra and neural network constructs.
Our focus was specifically on the best choice of high-level primitives, with an eye to compositionality and closure under transformations.
We achieved our aim by generalizing function composition, products and higher-order functions to an arbitrary number of arguments.
By using Naperian functors, we were able to take simple and obvious identities on functions and translate them for use with generic multidimensional array types, constructing subdivision and flip operations for higher-order functions operating on them.
As the driving goal of our work is to help develop tools which can automatically improve the hierarchical memory- and cache-efficiency as well as the parallelism of user-written programs, these transformations are derived entirely from the structure of the expression tree representing the computation written by the user.
Thus, this approach might be called structure-induced parallelism.
The rewrite rules we have expressed here are scalable to structures which are nested and subdivided to an arbitrary extent, and thus are potentially capable of distributing computations over the entire hierarchy of modern hardware, from vector instructions to entire clusters.
On the example of the matrix multiplication problem, we demonstrated how the automatic subdivision and reordering of higher-order functions can improve data locality by transforming an initial naive implementation.
 

\paragraph{Future work}
The approach to enumeration presented in this paper is limited to linear nesting of higher-order functions. In order for a generalization of this to enumeration of trees to be useful in practice, an early cut rule is also necessary to prune rearrangements which are not feasible, or which are not significantly different from previously enumerated cases. The enumerator as well as the early cut rule may depend on considerations which are problem- or platform specific; however, a solution can be based on top of the building blocks we have presented here.

There is an important set of problems not yet covered in the presented HoF formulation: the so-called sliding window problems, such as convolutions, finite difference schemes, and stencil operations on images.
The interaction of these with products, compositions, and reductions, and their rearrangement and subdivision, still needs to be addressed.




\section*{Acknowledgments}
D. B. is supported by the Hungarian National Bureau for Research, Development and Innovation, NKFIH No. K120660 and K123815. A. L. is supported by the \'UNKP-17-2 New National Excellence Program of the Ministry of Human Capacities. The authors are also thankful for the Wigner GPU Lab, especially M\'at\'e Ferenc Nagy-Egri and Bal\'azs Kacskovics for IT support and access to computation resources.








\end{document}